\documentclass{aa}  

\usepackage{natbib}
\usepackage{url}
\usepackage{graphicx}
\usepackage{txfonts}
\usepackage[]{hyperref}
\begin{document} 

\def\teff{$T\rm_{eff }$\,}
\def\kms{$\mathrm {km~s}^{-1}$}
\def\gtsima
{\hbox{\raise0.5ex\hbox{$>\lower1.06ex\hbox{$\kern-1.07em{\sim}$}$}}}
\def\ltsima
{\hbox{\raise0.5ex\hbox{$<\lower1.06ex\hbox{$\kern-1.07em{\sim}$}$}}}

\title{Chemical abundance analysis of red giant branch stars \\ in the
globular cluster E3\thanks{Based on observations made with ESO  Telescopes at
the La Silla Paranal Observatory under programme ID 097.D-0056(A).}}

   \author{L. \,Monaco\inst{1},
S. \,Villanova \inst{2},
G. \,Carraro\inst{3}, 
A. \,Mucciarelli\inst{4,5}
\and 
C. \,Moni Bidin\inst{6}
          }

\institute{
Departamento de Ciencias Fisicas, Universidad Andres Bello, Fernandez Concha 700, Las Condes, Santiago, Chile; \email{lorenzo.monaco@unab.cl}
\and
Universidad de Concepci\'on, Casilla 160-C, Concepci\'on, Chile
\and
Dipartimento di Fisica e Astronomia Galileo Galilei, Vicolo Osservatorio 3, I-35122, Padova, Italy
\and
Dipartimento di Fisica e Astronomia, Universit\`a degli Studi di Bologna, via Gobetti 93/2, I-40129 Bologna, Italy
\and
INAF - Osservatorio di astrofisica e scienza dello spazio di Bologna, Via Gobetti 93/3, I-40129, Bologna, Italy
\and
Instituto de Astronomia, Universidad Catolica del Norte, Av. Angamos 0610 Antofagasta, Chile
}
\authorrunning{Monaco et al.}

\titlerunning{E3 chemical abundances}

   \date{Received ...; Accepted...}

 
  \abstract
   {Globular clusters are known to host multiple stellar populations, which are
   a signature of their formation process. The globular cluster E3 is one of the
   few low-mass globulars that is thought not to host multiple populations.}
   {We investigate red giant branch stars in E3 with the aim of providing a
   first detailed chemical inventory for this cluster, we determine its radial
   velocity, and we provide additional insights into the  possible presence of
   multiple populations in this cluster.}
   {We obtained high-resolution FLAMES-UVES/VLT spectra of four red giant
   branch  stars likely members of E3. We performed a local thermodynamic
   equilibrium abundance  analysis based on one-dimensional plane parallel
   ATLAS9 model atmospheres. Abundances were derived from line equivalent widths
   or spectrum synthesis.}
   {We measured abundances of Na and of iron peak (Fe, V, Cr, Ni, Mn), $\alpha$ 
   (Mg, Si, Ca, Ti), and neutron capture elements (Y, Ba, Eu). The mean cluster
   heliocentric radial velocity, metallicity, and sodium abundance ratio are 
   v$_{helio}$=12.6$\pm$0.4\kms ($\sigma$=0.6$\pm$0.2\,\kms),
   [Fe/H]=-0.89$\pm$0.08\,dex, and [Na/Fe]=0.18$\pm$0.07\,dex, respectively. The
   low Na abundance with no appreciable spread is suggestive of a cluster
   dominated by first-generation stars in agreement with results based on lower
   resolution spectroscopy. The low number of stars observed does not allow us
   to  rule out a minor population of second-generation stars. The observed
   chemical abundances are compatible with the trends observed in Milky Way
   stars.}
   {}


   \maketitle
%

\section{Introduction}

Once considered a prototype of simple stellar populations \citep[][]{renzini88},
globular clusters (GCs) are now known to host multiple stellar populations
(MPs). First discovered in red giant branch   \citep[RGB, e.g.,][]{cohen78}
stars, chemical inhomogeneities in light elements persist down to the cluster
main sequence  \citep[MS,][]{gratton01} stars. The low central temperatures and
thin convective envelopes of these low-mass stars make them unable to be the
source of the observed abundance variations, which were then recognized as
signatures of the GCs formation process.

Globular clusters usually host at least two stellar populations, one with a
composition compatible with halo field stars (first generation, FG) and the
other with an enriched or polluted  composition (second generation, SG). The
observed abundance spread, most notably the Na-O
\citep[][]{carretta09a,carretta09b} and C-N \citep[][]{pancino10}
anti-correlations, accompanied sometimes by Mg-Al anti-correlations
\citep[][]{pancino17} and He abundance variations \citep[][]{pasquini11}, are
suggestive of hot hydrogen burning. One of the main questions yet to be answered
is the nature of the polluters which gave rise to enriched stars. Several models
have been proposed including pollution from massive asymptotic giant branch
stars \citep[][]{dercole16} and fast rotating massive stars
\citep[][]{decressin07}, among  others \citep[][]{bastian17}. However, none of
the models is fully satisfactory \citep[][]{renzini15} and in some cases,
different polluters may also be required \citep[][]{carretta18}.

Most GCs present a Na-O anti-correlation and are dominated by SG stars. The
fraction of stars belonging to the FG increases, however, with decreasing 
cluster mass \citep[][]{milone17}. The most massive open clusters and the least
massive globular clusters were both investigated in order to find empirical
evidence about the mass limit for the formation of multiple populations
\citep[][]{bragaglia12}. A few low-mass GCs do not present evidence of multiple
populations  \citep[][]{cohen04,sbordone07}, the most massive  being Rup\,106
\citep[][]{villanova13}. Nonetheless, MPs are detected in the less massive GCs
NGC\,6362, NGC\,6535, and ESO452-SC11 
\citep[][]{dalessandro14,mucciarelli16,bragaglia17,simpson17}. In addition to
mass, the age and metallicity may be two additional relevant parameters
\citep[][]{carretta10}. In particular, massive clusters older than  $\sim$2\,Gyr
are observed to host MPs, while younger ones are not \citep[see][and references
therein]{martocchia17,martocchia18,bastian17}.

With an absolute total magnitude of M$_V$=-4.12, E3 \citep[$\alpha$, $\delta$ =
09:20:57.07, --77:16:54.8;][2010 edition]{harris96} is one of  the faintest
globular clusters in the Galaxy and likely one of the least massive 
\citep[1.4$\times$10$^4$M$_\odot$;][]{salinas15}. Its color magnitude diagram,
with all the evolutionary phases beyond the MS severely contaminated by field
stars, as expected from its location (l = 292$^\circ$.270, b =
--19$^\circ$.020), made this cluster a particularly difficult one to be studied.

Recently, two low-resolution spectroscopy studies on E3 have been published:
\citet[hereafter SS15]{salinas15} and \citet[hereafter FM15]{delafuente15}. SS15
found no evidence for multiple stellar populations in this cluster from the
study of the strengths of the CH-CN bands of low-resolution spectra of 23 red
giant branch stars. FM15  have analyzed low-resolution, medium signal-to-noise
ratio (S/N) spectra of nine stars, finding two probable members and have derived
tentative radial velocity (RV) and metallicity estimates of 45$\pm$5\,\kms and
[Fe/H]=-0.7 dex. SS15, instead, have derived a RV of 8.9$\pm$2.8\kms. Both of
these RV determinations are based on just two stars and  are obviously in
conflict with each other. Robust determinations of its metallicity, RV, and
tangential velocity are essential to derive the cluster actual location and
motion and possibly associate E3 with other clusters (FM15).

We present here the first chemical abundance analysis of a sample of RGB stars
belonging to E3 based on high-resolution spectroscopy. In section \S\ref{obs} we
present the observations and data reduction. Sections \S\ref{aban} and
\S\ref{dis} present the performed abundance analysis and the obtained results.
Finally, in \S\ref{con} we present our conclusions.

\section{Observations and data reduction}\label{obs}

Observations were conducted using the multi-object, fiber fed FLAMES facility
\citep[][]{pasquini02} mounted at the unit two telescope (UT2,  ``Kueyen'') of
the European Southern Observatory (ESO) Very Large Telescope (VLT) located in
the Chilean Andes. FLAMES allows the simultaneous observation of up to seven or
eight  objects (depending on the plate used for the observations) using the red
arm of the high-resolution  UV-Visual Echelle Spectrograph
\citep[UVES,][]{dekker00}. FLAMES-UVES observations deliver spectra with a
resolution of R$\simeq$47,000. 

We observed seven stars using FLAMES-UVES, set at central wavelength 580\,nm,
which covers the spectral range 476--684\,nm, with a 5\,nm gap around the
central wavelength. One fiber was placed on an empty field for the purpose of
sky subtraction. Three 3000\,s exposures were taken in service mode on the
nights of April 25, May 19, and May 21, 2016.  

\begin{figure}  
\includegraphics[width=1\columnwidth]{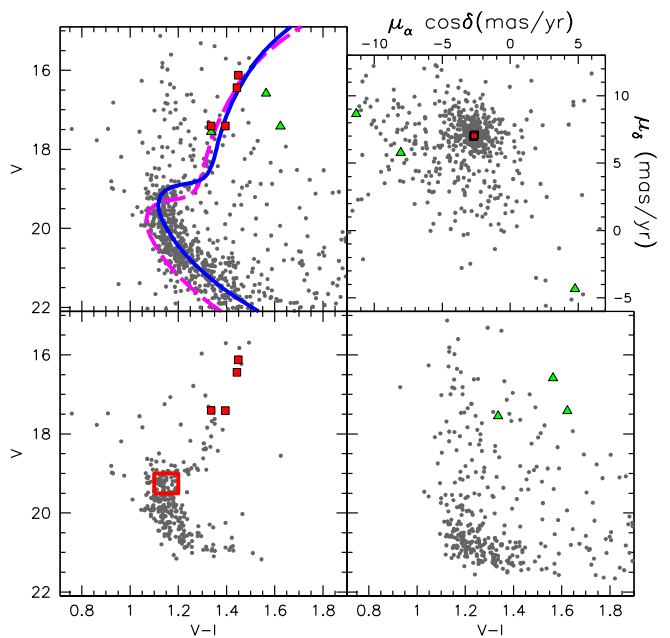}  

\caption{Upper left panel: E3 V {vs} V-I color magnitude diagram. Target stars
are marked by large filled symbols. Filled red squares are E3 radial velocity
members. The continuous and dashed lines are Z=0.003, 13\,Gyr isochrones from
the PARSEC collection \citep[][]{bressan12} where the visual distance modulus
and reddening are the values we adopt here from FM15 (continuous blue line) or
from  \citet[][2010 revision, dashed magenta line]{harris96}. Upper right panel:
Gaia DR2 proper motions of stars cross-identified from the photometry in the
upper left panel. RV members (red filled squares) have very similar PMs,  and 
thus are  almost superposed on each other in the figure. Lower left panel: CMD
of stars having PMs within three times the errors from the mean E3 PM. Stars in
the cross-identified  catalog and having 19.0$<$V$<$19.5; 1.1$<$V-I$<$1.2 (red
box) were used to select the MS stars that were used to define the cluster mean
PM.	 Lower right panel: CMD of stars having PMs exceeding three times the
errors from the mean E3 proper motion.}\label{cmd}  

\end{figure}

Table\,\ref{PA} lists the target stars identification numbers, coordinates, and
V- and I-band photometry. Only two 3000\,s observations were obtained for object
\#132. The seven stars observed are plotted as filled symbols in the V versus
V-I E3 color magnitude diagram (CMD) in Figure\,\ref{cmd}  (upper left panel). 
We adopt here the photometry described in FM15. Star \#100 excluded, our targets
are all present in the \citet[][hereafter V96]{veronesi96} photometry.  For
comparison, the color difference between the V96 V-I and ours are below
0.01\,mag for all stars but \#1. In this case the V96 V-I color is 0.028\,mag
smaller than ours.  This  difference corresponds to a temperature about 50\,K
hotter when using the color to get the stellar effective temperature, well below
the typical temperature errors assumed in chemical abundance analysis. A 13\,Gyr
isochrone of metallicity Z=0.003 from the PARSEC
collection\footnote{\url{http://stev.oapd.inaf.it/cgi-bin/cmd\_2.8}} 
\citep[][]{bressan12} was superposed to the cluster CMD, adopting E(V-I)=0.47
and (m-M)$_V$=15.07 from FM15 (our favored values, continuous blue line) and 
E(V-I)=0.42 and (m-M)$_V$=15.47 from \citet[][2010 edition, dashed magenta
line]{harris96}. The \citet[][]{schlegel98} reddening maps suggest that E3 is
not significantly affected by differential reddening. The mean reddening  at the
positions of the seven stars we observed is E(B-V)=0.340$\pm$0.005 or
E(B-V)=0.293$\pm$0.004, according to \citet[][]{sf11} with the maximum variation
among the positions on the order of 0.01\,mag.  We note  that
\citet[][]{sarajedini07} suggest for E3 an age that is  $\sim$2\,Gyr younger
than 47\,Tuc. As can be seen from Fig. 3 of \citet[][]{delafuente15}, at these
old ages, differences of a few Gyr are fully compatible with our photometry. On
the other hand, \citet[][]{marinfranch09} classify E3 among the group of ``old''
globular clusters (12.8$\pm$1.4\,Gyr in their D07 scale) and did not find a
significantly younger age with respect to 47\,Tuc (13.1$\pm$0.9\,Gyr in the same
scale).

\begin{table*}

\caption{Target IDs, coordinates, photometry, atmospheric parameters,  measured
metallicities, spectral S/N, radial velocities and Gaia DR2 proper motions.}

\label{PA}
\small
\begin{center}
\begin{tabular}{lcccccccccrrrr}
\hline
ID  & $\alpha$(J2000) & $\delta$(J2000)          &    V  &    V-I & \teff         &log g & $\xi$  & [Fe/H] & S/N & v$_\mathrm{helio}$ & $\mu_\alpha cos\delta$ & $\mu_\delta$\\\
   &&&&  & K             &      & \kms    &       &  @600nm &\kms & mas/yr & mas/yr\\
\hline
\object{1}         & 09:20:50.59   & -77:19:55.7   & 17.55  &  1.34      &       &      &     &      &   8  &  -0.2$\pm$0.6 & -11.32$\pm$0.11 &  8.66$\pm$0.12\\
\object{67}        & 09:21:15.65   & -77:18:04.0   & 16.44  &  1.44      &4807  &2.53  &1.29 &-0.89 &  14  &  11.8$\pm$1.0 &  -2.69$\pm$0.06 &  6.95$\pm$0.06\\ 
\object{99}        & 09:21:01.01   & -77:16:34.3   & 17.41  &  1.34      &5047  &3.04  &1.16 &-0.82 &  10  &  13.3$\pm$0.1 &  -2.62$\pm$0.11 &  7.09$\pm$0.12\\ 
\object{100}       & 09:21:50.61   & -77:16:34.1   & 17.41  &  1.39      &4923  &2.98  &1.17 &-0.84 &   8  &  12.8$\pm$1.2 &  -2.62$\pm$0.11 &  7.08$\pm$0.11\\ 
\object{107}       & 09:20:16.73   & -77:15:56.9   & 16.12  &  1.45      &4785  &2.39  &1.32 &-1.01 &  24  &  12.3$\pm$0.8 &  -2.67$\pm$0.06 &  7.07$\pm$0.05\\
\object{121}       & 09:20:21.89   & -77:18:47.8   & 16.58  &  1.56      &       &      &     &      &  11  &  39.1$\pm$3.9 &   4.78$\pm$0.07 & -4.34$\pm$0.07\\
\object{132}       & 09:20:45.94   & -77:15:12.8   & 17.42  &  1.62      &       &      &     &      &   8  &  29.0$\pm$0.7 &  -8.05$\pm$0.12 &  5.77$\pm$0.13\\
                                                
\hline
\end{tabular}
\end{center}
\normalsize
\end{table*}                              

Raw science data were retrieved through the ESO data
archive\footnote{\url{http://archive.eso.org/}} together with the associated
master calibrations as delivered by the system. Raw data were then reduced using
those calibrations and the FLAMES-UVES CLP based pipeline version
5.5.5\footnote{\url{http://www.eso.org/sci/software/pipelines/}}. The
``flames\_obs\_scired'' recipe alone was applied. For each exposure, the
spectrum corresponding to the sky position was subtracted from the individual
stellar spectra. 

\begin{figure}
\includegraphics[width=1\columnwidth]{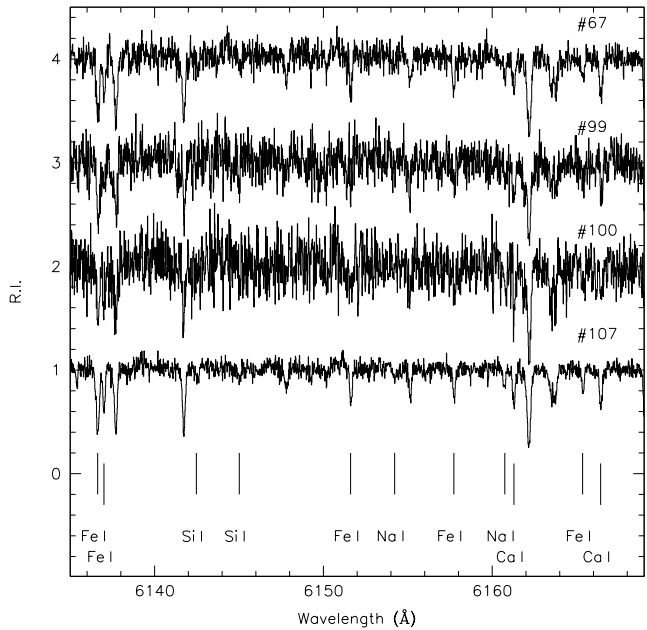}
\caption{Sample of the spectra of the observed stars. A few absorption lines 
of different elements are indicated.}\label{spec}
\end{figure}

For each star and epoch, the RV was measured using the {\tt fxcor} task within
IRAF\footnote{IRAF is distributed by the National Optical Astronomy Observatory,
which is operated by the Association of Universities for Research in Astronomy
(AURA) under a cooperative agreement with the National Science Foundation.} to
cross-correlate the observed spectra with a template synthetic spectrum selected
from the \citet[][]{coelho05} collection. Heliocentric corrections were
calculated using the {\tt rvcorrect} task in IRAF. Finally, individual stellar
spectra were corrected to rest frame and median combined. The signal-to-noise
ratios measured on the combined spectra at about 600nm are also given in
Table\,\ref{PA} together with the mean  heliocentric radial velocity. The errors
reported on the RVs are the standard deviation of the available measurements.
Figure \ref{spec} presents a sample of the obtained spectra, with a few
absorption lines marked for reference.

\begin{figure}
\includegraphics[width=1\columnwidth]{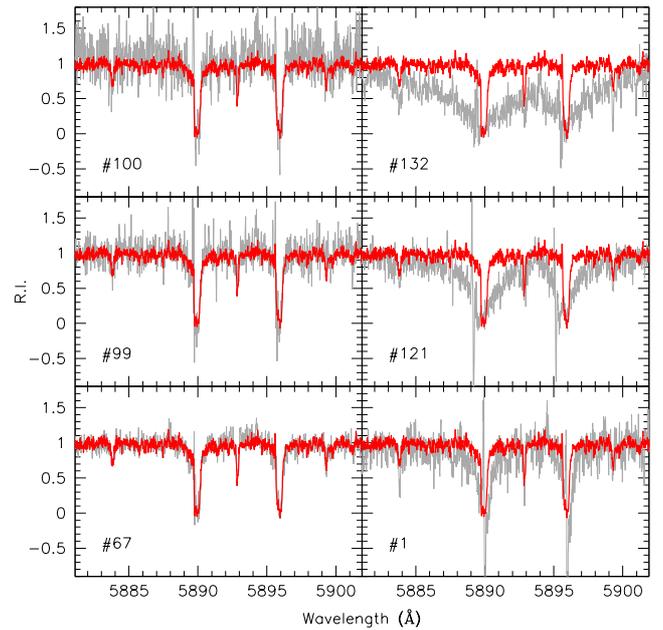}
\caption{Sample of the spectra of the observed stars in the region of the Na\,D doublet.
The spectrum of star \#107 is marked in red and superposed on the spectra of
the other stars (gray). The spectra of stars \#1, \#121, and \#132 present 
significantly broader Na\,D lines than stars \#67, \#99, \#100, and \#107.}
\label{na}
\end{figure}

Stars \#67, \#99, \#100, and \#107 have very similar RVs, significantly
different from the remaining three stars. The locations of these four stars in
the CMD in Figure\,\ref{cmd} (upper left panel, red filled squares) are
compatible with the expected location of the cluster RGB (continuous blue line).
The position of star \#1 is also   compatible with the cluster RGB. Its RV is,
however, significantly different from that of the other four stars. We show in
Figure\,\ref{na}, a portion of the stellar spectra, centered around the sodium D
doublet (Na\,D). The spectrum of star \#107 (thick continuous red line) is
superposed on the spectra of the stars indicated in each subpanel (continuous
gray lines). The shape of the Na\,D lines is very similar for stars \#67, \#99,
\#100, and \#107 (left panels). On the other hand, the spectrum of star \#1  has
significantly broader Na\,D lines (bottom right panel) even though it has the
same color  as star \#99. This suggests that the observed line broadening is a
pressure effect, and that star \#1 is a dwarf star rather than a giant. The
spectra of stars \#121 and \#132 present very broad Na\,D lines (top and middle
right panels). This is expected, however, given their redder color and cooler
temperatures with respect to the remaining stars. Pressure effects are 
therefore  not easily disentangled. 

We cross-identified stars in our reference photometry with the Gaia DR2 catalog
\citep[][]{gaia16,gaia18} using the {\tt TOPCAT} code \citep[][]{topcat}. The
upper right panel of Fig.\,\ref{cmd} shows the  proper motions (PM) measured by
the Gaia mission;  our target stars are identified with the same symbols as in
upper left panel. Stars  \#67, \#99, \#100, and \#107 (RV members, red filled
squares) also present  very similar PMs, while stars \#1, \#121, and \#132 have
significantly scattered values. Using our photometry, we selected E3 main
sequence stars using the box shown for reference in the lower left panel
(19.0$<$V$<$19.5; 1.1$<$V-I$<$1.2). We also required errors in PM lower than 0.5
mas/yr and differences between the V and g magnitudes in the range from -0.35 to
-0.10, according to the corresponding distributions. We further excluded one
outlier from the resulting PM distributions. This way, we selected 44 likely
cluster member MS stars, from which we obtain a mean E3 PM of $(\mu_\alpha
cos\delta; \mu_\delta) = (-2.71\pm0.40; 7.20\pm0.46)$ mas/yr. The lower panels
of Fig.\,\ref{cmd} present the CMDs of stars having PMs within (left) or
exceeding (right) three times the quoted errors from the mean cluster PM. The
Gaia PMs allow us to nicely distinguish the E3 cluster population, clearly
visible in the lower left panel, from the contaminating field stars (lower right
panel). We note  that the four RV members have PMs fully compatible with the
cluster PM derived from MS stars (lower left and upper right panels, see also
Table\,\ref{PA}). We therefore consider that   stars \#1, \#121, and \#132  are
not cluster members.

Considering only stars \#67, \#99, \#100, and \#107  as members, we obtain a
mean cluster radial velocity and dispersion of v$_{helio}$=12.6$\pm$0.4\,\kms
and $\sigma$=0.6$\pm$0.2\,\kms, where the errors were in both cases evaluated
through a jackknife resampling technique \citep[][]{lupton93}. The cluster mean
RV we obtained is in reasonable agreement with SS15, who measured
8.9$\pm$2.8\,\kms\ from intermediate-resolution (0.75\AA) optical spectroscopy
($\sim$5100--5600\AA)  of two stars. On the other hand, FM15 measured
45$\pm$5\kms, which suggests the two stars they measured to be likely
non-cluster members. Even though it is among the clusters with the largest
binary fraction \citep[][]{veronesi96,milone12} and within the limitation of our
reduced statistic, the E3 velocity dispersion is the smallest found in a
globular cluster, after Pal\,14 (0.4$\pm$0.1\kms), according to the
\citet[][2010 edition]{harris96} catalog. This low $\sigma$ supports a very low
present day mass for E3, in agreement with the suggestion of SS15, who estimates
a cluster mass of 1.4$\times$10$^4$M$_\odot$. 

\section{Abundance analysis}\label{aban}

\begin{table*}

\caption{Target stars chemical abundances. The penultimate column lists average
abundances and Gaussian dispersions. The adopted solar abundances are given in
the last column. Numbers in parentheses indicate the error of the mean and the
number of lines used. An ``s''  character after the second number in parentheses
is used to indicate a measure performed through spectrosynthesis. Star \#100 was
excluded from the computation of the mean iron  abundance and dispersion from
Fe\,II lines.}\label{Ab}

\begin{center}
\begin{tabular}{l|l|l|l|l|c|c}
\hline
           Element               &  \#67                &       \#99             &        \#100                  &        \#107                  & $<$[X/Y]$>\pm\sigma$ &       A(X)$_\odot$  \\
\hline 
         {[}FeI/H]               &  -0.89 (0.01/56)     &    -0.82 (0.03/42)    &  -0.84        (0.03/51)    &   -1.01       (0.01/63)       & -0.89$\pm$0.08  &      7.50    \\      
         {[}FeII/H]              &  -0.85 (0.05/5)      &    -0.74 (0.05/4)     &  -0.43        (0.08/5)    &   -1.04       (0.03/5)        & -0.88$\pm$0.15  &      7.50    \\      
         {[}Na/Fe]               &   0.12 (-/1)         &     0.17 (0.01/2)     &   0.29        (-/1)       &    0.16       (0.02/2)        &  0.18$\pm$0.07  &      6.37    \\     
         {[}Mg/Fe]               &   0.13 (-/1s)        &     ---                &   ---                         &    0.43       (-/1s)          &  0.28$\pm$0.21  &     7.54      \\     
         {[}Si/Fe]               &   0.38 (0.06/2)      &     0.47 (0.07/2)     &   0.42  (0.01/2)        &    0.38       (0.05/4)        &  0.41$\pm$0.04  &      7.61    \\     
         {[}Ca/Fe]               &   0.37 (0.03/4)      &     0.36 (0.08/4)     &   0.27  (0.11/4)        &    0.36       (0.02/5)        &  0.34$\pm$0.05  &      6.39    \\    
         {[}Ti/Fe]               &   0.32 (0.02/5)      &     0.29 (0.05/3)         &   0.21        (-/1)           &    0.38       (0.03/8)        &  0.30$\pm$0.07  &       4.94    \\   
         {[}V/Fe]                &   0.07 (-/1s)        &     ---               &    ---                  &    0.22       (-/1s)          &  0.14$\pm$0.11  &      4.00    \\
         {[}Cr/Fe]               &   0.20 (-/1)         &     ---               &   0.02  (-/1)           &    0.10       (-/1)           &  0.11$\pm$0.09  &      5.63    \\   
         {[}Ni/Fe]               &   0.05 (0.02/10)     &     0.03 (0.08/5)     &  -0.04  (0.09/6)        &    0.02       (0.02/8)        &  0.01$\pm$0.04  &      6.26    \\   
         {[}Mn/Fe]               &   ---                &     ---               &   ---                   &   -0.22       (-/1s)          &                 &     5.37     \\     
         {[}Y/Fe]$_{509nm}$      &   ---                &     ---               &   ---                   &   -0.08       (-/1s)          &                 &     2.25     \\ 
         {[}Y/Fe]$_{540nm}$      &   ---                &     ---               &   ---                   &    0.19       (-/1s)          &                 &     2.25     \\ 
         {[}Ba/Fe]               &   0.01 (-/1s)        &    -0.06 (-/1s)       &   0.11  (-/1s)          &    0.07       (-/1s)          &  0.03$\pm$0.07  &      2.34    \\  
         {[}Eu/Fe]               &   ---                &     ---               &   ---                          &    0.49       (-/1s)          &                   &     0.52    \\   
\hline
\end{tabular}
\end{center}
\end{table*}

Chemical abundance analysis was performed on the spectra of stars \#67, \#99
\#100, and \#107 using the local thermodynamic equilibrium code
MOOG\footnote{\url{http://www.as.utexas.edu/~chris/moog.html}}
\citep[][]{sneden73}. Appropriate one-dimensional ATLAS\,9 model atmospheres
were calculated \citep[][]{k93,sbordone04} for the analysis. 

Stellar effective temperatures were derived from the V-I colors, using the
\citet[][]{alonso99} calibrations ($\sigma$(\teff)=125\,K). The V-I colors were
converted to the Johnson system using the relation provided by
\citet[][]{bessell83}. We adopted E(V-I)=0.47 from FM15. Had we used a reddening
of E(B-V)=0.30\,mag or E(V-I)=0.42 from \citet[][]{sf11}, our temperatures would
have resulted between 108--124\,K colder (average 114\,K). A reddening of
E(B-V)=0.25 mag or E(V-I)=0.35 from the \citet[][]{schlegel98} reddening maps
corrected according to \citet[][]{bonifacio00} would correspond instead to
$\sim$260K colder temperatures ($\Delta$\teff=262$\pm$18\,K). Surface gravities
were obtained from the standard relation 

$$
log\,g=log\frac{M}{M_\odot}+4\,log\frac{T_{eff}}{5777}+0.4(M_{bol}-4.76)+4.44~,
$$

\noindent where we made use of the solar effective temperature, surface gravity,
and bolometric absolute magnitudes of 5777\,K, 4.44\,dex, and 4.76\,mag,
respectively. The bolometric magnitudes were calculated from the visual values
using again the apparent visual distance modulus from FM15, i.e.,
(m-M)$_V$=15.07. The bolometric correction was obtained from the
\citet[][]{alonso99} calibration for the appropriate \teff and metallicity,
along with a stellar mass of 0.8\,M$_\odot$. A variation of 150\,K in \teff
results in a change of about 0.07\,dex in log\,g.

The distance modulus we employed is 0.4\,mag smaller than the value reported by
\citet[][]{harris96}, namely (m-M)$_V$=15.47. A Z=0.003, 13 Gyr old isochrone
would not be a reasonable match to the cluster main sequence, if we used
(m-M)$_V$=15.47 and E(V-I)=0.42 (dashed magenta line in Fig.\,\ref{cmd},  upper
left panel) or  E(V-I)=0.35.  We note  that \citet[][]{sarajedini07} obtained,
from the analysis of their HST CMD, the same reddening value we adopt here. If
we adopted this latter distance modulus, we would obtain surface gravities
0.16\,dex lower.  Finally,  microturbulent velocities were obtained from the
\citet[][]{marino08} calibration: $\xi$=-0.254\,log\,g+1.930. A variation of
0.2\,dex in log\,g would cause a change of 0.05\,\kms in $\xi$.

In the analysis, we employed the same line list we used in other publications of
our group \citep[see, e.g.,][]{villanova16} and we refer the reader to
\citet[][]{villanova11} for details. Abundances were derived from line  
equivalent widths (EWs) for Fe, Na, Si, Ca, Ti, Cr, and Ni and through
spectrosynthesis for Mg, V, Mn, Y, Ba, and Eu.

The measured iron abundances are listed in Tables\,\ref{PA} and \ref{Ab}.
Table\,\ref{Ab} also gives the species over iron abundance ratios, the adopted
solar abundances, and the mean species abundance and Gaussian dispersion for the
sample when the abundance was measured for more than one star. For each star and
species we give  the error of the mean (when more than one line was used) and
the number of lines used. An ``s'' indicates whether the line was measured via
spectrosynthesis.

Following \citet[][]{cayrel88}, given the resolution and sampling of our
spectra, we expect errors on the measured EWs to be on the order of
$\sim$4--11\,m\AA, depending on the spectral S/N. We considered a line as
detected if its EW exceeds the corresponding uncertainty by at least three
times.   Program stars have similar atmospheric parameters (see
Table\,\ref{PA}). Table\,\ref{errors} presents the abundance changes resulting
for variations in the atmospheric parameters of $\Delta$\teff=$\pm$150\,K,
$\Delta$log\,g=0.2\,dex, and $\Delta\xi$=$\pm$0.05\kms, taken as representative
of the internal uncertainties. We note, however, that temperatures 260\,K colder
than those we adopted would correspond to about 0.28\,dex lower metallicities.
We also allow for an  uncertainty in the measured iron abundance of
$\pm$0.15\,dex. When the abundance was derived from at least three lines, we
consider the error of the mean listed in Table\,\ref{Ab} as representative of
the error induced by the spectral S/N. For elements derived from one or two
lines only, this uncertainty is set  to the error returned by the fitting
procedure when the abundance was obtained through spectrosynthesis, or to the
abundance uncertainty induced by a variation in EW equal to the values mentioned
above, following \citet[][]{cayrel88}. We list under column $\epsilon$(S/N) the
minimum and maximum observed values. Given the limited and different quality of
the stellar spectra under analysis, in order to perform an analysis that is as
homogeneous as possible for the sample stars, we decided to keep the \teff,
log\,g, and $\xi$ fixed at their initial values.

For the calculation of the abundance variations in Table\,\ref{errors}, we
adopted the atmospheric parameters and abundances of star \#107. Uncertainties
in the \teff, log\,g, and $\xi$ are correlated and are summed directly. This is
then summed in quadrature to the uncertainties in A(Fe) and from the spectral
S/N. The penultimate column lists the overall abundance uncertainty for each
species. Two values are provided corresponding to the minimum and maximum S/N
error given in the previous column, when applicable. Finally, in the last column
we list the observed Gaussian dispersion in the measured abundances when
measures were performed for more than one star.

The iron abundances obtained from the Fe\,I and Fe\,II lines are in good
agreement (differences lower than 0.1\,dex) for stars \#67, \#100, and \#107,
lending support to the adopted gravities. For star \#100, on the other hand, we
find a 0.4\,dex abundance difference. The higher iron abundance derived from
Fe\,II lines may originate from an incorrect placement of the continuum in the
low S/N (the lowest in the sample) spectrum of star \#100 which might have lead
to overestimate the EWs of the weak Fe\,II lines. Differences between the iron
abundance derived from Fe\,I and Fe\,II lines have been reported for a few
clusters that are thought  to present an iron spread \citep[see][and references
therein]{mucciarelli18}. In those cases, however, the differences were smaller
and iron abundances based on Fe\,II lines are considered to be more reliable. In
the case of star \#100, this seems unlikely, and the iron content derived from
Fe\,I lines should be considered more robust, due to the larger number of
measured lines and the agreement with the iron abundance derived for the other
stars in the sample from both Fe\,I and Fe\,II lines. A log\,g 0.4\,dex lower
for this star would decrease the A(FeII) and the A(Ba) abundances by about 0.18
and 0.12\,dex, respectively, bringing in better agreement the iron abundances
derived from Fe\,I and Fe\,II lines and the Ba abundance of this star with the
values derived for the other stars. The mean iron abundance and dispersion from
Fe\,II lines given in Tables\,\ref{Ab} and \ref{errors} were computed excluding
star \#100. With the inclusion of this star we would obtain instead
[Fe\,II/H]=-0.76$\pm$0.26\,dex.

\section{Results}\label{dis}

\begin{table*}

\caption{Sensitivities of the derived abundances to the indicated variation of
the atmospheric parameters. The last two columns indicate the combined total
expected uncertainty and the observed dispersion from Table\,\ref{Ab},
respectively. Column $\epsilon$(S/N) indicates the minimum and maximum abundance
variation observed from Table\,\ref{Ab} (error of the mean) or, when only  one
or two lines  were measured, expected according to the EW error induced from the
spectral S/N or as returned from the fitting procedure in the case of spectrum
synthesis.}\label{errors}

\begin{center}
\begin{tabular}{c|c|c|c|c|c|c|c}
\hline
&  $\Delta$\teff               &        $\Delta$log\,g                 &        $\Delta \xi$    &  $\Delta$A(Fe)     &       $\epsilon$(S/N)  &   Total  &   Observed   \\
&  $\pm$150\,K                 &        $\pm$0.2\,dex                  &        $\pm$0.05\kms   &  $\pm$0.15          &                   &          &        \\ 
\hline 

{[}Fe\,I/H]             & 0.16 & 0.00 & 0.02 & 0.00 & 0.01 / 0.03      & 0.18                  &  0.08\\
{[}Fe\,II/H]            & 0.06 & 0.09 & 0.01 & 0.00 & 0.03 / 0.08      & 0.16 / 0.18           &  0.15\\
{[}Na\,I/Fe]            & 0.05 & 0.00 & 0.02 & 0.01 & 0.07 / 0.18      & 0.10 / 0.19          &  0.07\\
{[}Mg\,I/Fe]            & 0.03 & 0.03 & 0.01 & 0.00 & 0.08 / 0.12      & 0.11 / 0.14           &  0.21\\
{[}Si\,I/Fe]            & 0.14 & 0.03 & 0.01 & 0.02 & 0.05 / 0.21      & 0.19 / 0.28          &  0.04\\
{[}Ca\,I/Fe]            & 0.00 & 0.01 & 0.01 & 0.01 & 0.02 / 0.11      & 0.03 / 0.11           &  0.05\\
{[}Ti\,I/Fe]            & 0.05 & 0.00 & 0.01 & 0.01 & 0.02 / 0.20      & 0.06 / 0.21           &  0.07\\
{[}V\,I/Fe]             & 0.06 & 0.00 & 0.02 & 0.02 & 0.13             & 0.15                  &  0.11\\
{[}Cr\,I/Fe]            & 0.07 & 0.00 & 0.01 & 0.01 & 0.07 / 0.20      & 0.11 / 0.22           &  0.09\\
{[}Ni\,I/Fe]            & 0.02 & 0.02 & 0.01 & 0.01 & 0.02 / 0.09      & 0.05 / 0.10           &  0.04\\
{[}Mn\,I/Fe]            & 0.03 & 0.00 & 0.02 & 0.01 & 0.04             & 0.06                  &      \\   
{[}Y\,II/Fe]$_{509nm}$  & 0.11 & 0.06 & 0.00 & 0.04 & 0.12             & 0.21                  &      \\   
{[}Y\,II/Fe]$_{540nm}$  & 0.16 & 0.08 & 0.01 & 0.05 & 0.11             & 0.28                  &      \\   
{[}Ba\,II/Fe]           & 0.10 & 0.06 & 0.03 & 0.04 & 0.05 / 0.10      & 0.20 / 0.22           &  0.07\\   
{[}Eu\,II/Fe]           & 0.16 & 0.08 & 0.01 & 0.05 & 0.11             & 0.28                  &      \\   

\hline
\end{tabular}
\end{center}
\end{table*}                              

The observed species abundance spreads (last column in Table\,\ref{errors}) are
typically  on the same order of or smaller than the expected uncertainties
(penultimate column), due to the uncertainties in the atmospheric parameters and
the spectral S/N.

\begin{figure}
\includegraphics[width=1\columnwidth]{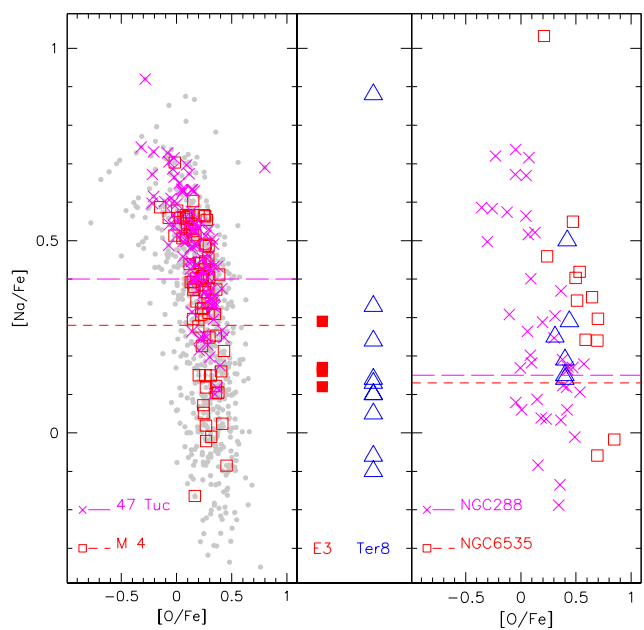}

\caption{Left panel: Na-O anti-correlation ([Na/Fe] {vs} [O/Fe]) for the sample
of GCs studied by \citet[][]{carretta09a,carretta09b}. M4 and 47\,Tuc stars are 
marked as red open squared and magenta crosses, respectively. The short-
long-dashed red and magenta lines indicate the separation between FG and SG
stars in the two clusters, according to \citet[][]{carretta10}. Right panel:
Same as left panel,  but for NGC\,288 (magenta crosses), Ter\,8 \citep[][open
blue triangles for stars with both O and Na measures]{carretta14}, and NGC\,6535
\citep[][upper limits excluded]{bragaglia17}. Middle panel: E3 (filled red
squares) and Ter\,8 \citep[][stars with Na measures only]{carretta14} Na
abundances.}\label{na2}

\end{figure}

We measured Na abundances between 0.12 and 0.29\,dex, i.e., a 0.17\,dex
variation and a Gaussian dispersion of 0.07\,dex, which is compatible with the
expected measurement uncertainties. The studied cluster stars thus have
homogeneous Na abundances within the allowed errors. Sodium abundances were
corrected for departures from local thermodynamic equilibrium following
\citet[][]{gratton99}. Unfortunately, in our spectra the 6300.3\AA\, oxygen line
is severely contaminated by a close, strong emission sky line. As a consequence,
we were not able to measure or put an upper limit to the O abundance of the
target stars.

In the left panel of Fig.\,\ref{na2} we present the Na-O anti-correlation for
the collection of GCs studied by \citet[][gray filled circles, upper limits
excluded]{carretta09a,carretta09b}. Data for the GCs M4 and 47\,Tuc having
metallicities close to E3 are also marked as open red squares and magenta
crosses, respectively. The short- and long-dashed  horizontal lines indicate the
separation between the first and second  stellar generation in these clusters,
as defined by \citet[][]{carretta10}. These limits are set 0.3\,dex above the
observed minimum Na abundance (or [Na/Fe]$_{min}$+4$\sigma$), excluding obvious
outliers. In the right panel we present the corresponding data for the
intermediate- and low-mass globular clusters NGC\,288
\citep[M$_V$=-6.75,][]{harris96}, Ter\,8 \citep[M$_V$=-5.07][stars with both O
and Na measures only]{carretta14} and NGC\,6535 \citep[M$_V$=-4.75][upper limits
excluded]{bragaglia17}.  Finally, in the middle panel we show the Na abundances
of E3 stars and the Ter\,8 stars from \citet[][]{carretta14} for which no O
abundance was measured. E3 stars are placed in the lower portion of the diagram,
which is populated by  stars less enriched in Na. All of our E3 stars are close
to or below  the line separating FG and SG stars in M4 and 47\,Tuc. Three out of
four E3 stars are also very close to the line dividing FG and SG stars in the
low- and intermediate-mass clusters NGC\,288 and NGC\,6535. In the case of
Ter\,8, \citet[][]{carretta14} tentatively identified as belonging to the SG
only the star having the highest Na abundance ratio  ([Na/Fe]=+0.88), and E3
stars present Na abundances similar to the remaining stars.

The stars we analyzed thus belong  to the FG, and we found no evidence of stars
belonging to the SG. E3 is therefore dominated by FG stars. According to
Fig.\,22 of \citet[][]{milone17}, the fraction of FG stars in E3 is indeed
expected to exceed 60\% (M$_V$=-4.12, 1.4$\times$10$^4$M$_\odot$). With our
limited statistics, which is based on only four stars, we cannot exclude  the
presence of SG stars. For example, \citet[][]{dalessandro16} observed five RGB
stars in NGC121, all of which belong to the FG, while from photometry they were
able to detect the presence of the SG and concluded that FG stars account for
more than 65\% of the total cluster mass; however,  in NGC121 SG  stars are more
centrally concentrated than FG stars.  This, combined with the target selection
from the outer cluster regions and the higher incidence of FG stars, favors the
bias toward the observation of the latter. Our results are also consistent and
support the results of  SS15, who analyzed the CH-CN band in low-resolution
spectra of 23 RGB stars and found no evidence of MPs in E3.

We measured Mg abundances in only two stars  (\#67 and \#107) by
spectrosynthesis of the strong, saturated line at 5711.083\,\AA. The [Mg/Fe]
abundance ratios in the two stars differ by 0.3\,dex, to be compared with
estimated uncertainty of 0.11--0.14\,dex. Dispersion in Mg abundances are
observed in GCs, and participate in the observed anti-correlations
\citep[e.g.,][]{pancino17}. Magnesium spreads are, however, usually less
pronounced than  Na spreads \citep[][]{pancino17}, whereas we did not observe
any spread in Na. We note that the difference in the EW of the two lines
($\sim$10\,m\AA) is compatible with the expected errors of 3.6 and 6.2\,m\AA\,
for stars \#107 and \#67, respectively, and that part of the 0.3\,dex difference
in the [Mg/Fe] abundance ratio originates from the difference in the measured
iron content. Given the uncertainties involved in the present analysis, the
observed variation in Mg abundance between stars \#67 and \#107 has to be
considered  as only tentative and wait for confirmations from additional
analysis.

\begin{figure}
\includegraphics[width=1\columnwidth]{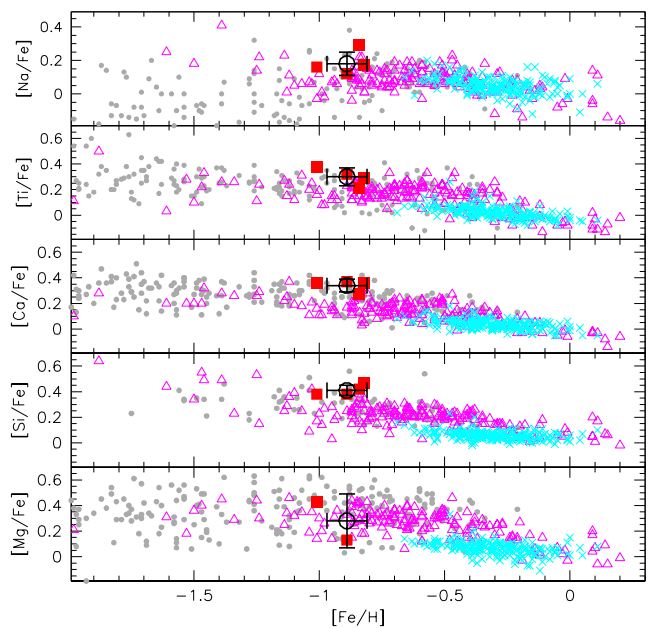}

\caption{Run of [Na/Fe] and $\alpha$-element abundance ratios ([Ti/Fe], [Ca/Fe],
[Si/Fe],  [Mg/Fe] from top  to bottom) {vs} [Fe/H] for E3 (filled red squares)
and Galactic stars. The large black open circle with error bars indicates the
mean cluster abundance and Gaussian dispersion. Gray filled circles are halo
stars from \citet[][their dissipative component]{gratton03} and \citet[][halo
probability greater than 0.8]{venn04}. Open magenta triangles are thick disk
stars from \citet[][]{reddy06}, while cyan crosses are thin disk stars from
\citet[][]{reddy03}.}\label{alpha}

\end{figure}

\begin{figure}
\includegraphics[width=1\columnwidth]{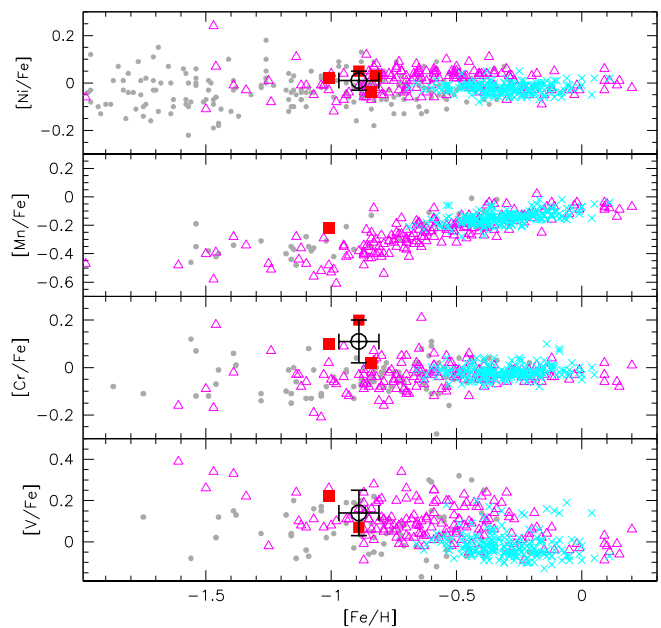}
\caption{Run of iron peak element abundance ratios ([Ni/Fe], [Mn/Fe], [Cr/Fe], 
[V/Fe] from top  to bottom) {vs} [Fe/H] for E3 and Galactic stars. Symbols
are the same as in Fig. \ref{alpha}.}\label{iron}
\end{figure}

\begin{figure}
\includegraphics[width=1\columnwidth]{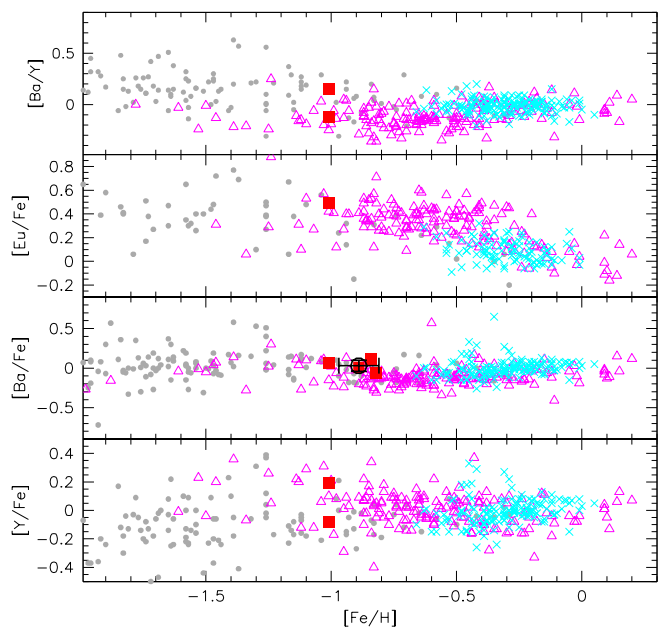}

\caption{Run of neutron capture element abundance ratios ([Ba/Y], [Eu/Fe],
[Ba/Fe], [Y/Fe] from top  to bottom) {vs} [Fe/H] for E3 and Galactic stars.
Symbols are the same as Fig.\ref{alpha}.}

\label{heavy}
\end{figure}

Figure \ref{alpha} shows the E3  $\alpha$-element abundance ratios ([Ti/Fe],
[Ca/Fe], [Si/Fe], and [Mg/Fe] filled squares) versus  [Fe/H] in comparison with
Galactic stars. In the figure, gray filled circles are halo stars from
\citet[][their dissipative component]{gratton03} and \citet[][halo probability
greater than 0.8]{venn04}. Open magenta triangles are thick disk stars from
\citet[][]{reddy06}, while cyan crosses are thin disk stars from
\citet[][]{reddy03}. The [Mg/Fe] abundance ratio of star \#67 appears low, but
it is within the observed trend. The mean cluster abundance ratios (large empty
circles with error bars) are, in all cases, well within the observed Galactic
trends, with enhancements consistently between $\sim$0.3 and 0.4\,dex. Agreement
with the Galactic trends are also observed for the abundance ratios over Fe of
Na  (upper panel in Fig.\,\ref{alpha}), iron-peak elements (V, Cr, Mn, and Ni,
Fig.\,\ref{iron}) and neutron capture elements (Y, Ba, and Eu,
Fig.\,\ref{heavy}).

We measured the yttrium abundance of star \#107 only by spectrosynthesis of two
lines. The abundances of the two lines differ by about 0.3\,dex (-0.08 and
0.19\,dex) and we provide the values for the individual lines in
Table\,\ref{Ab}. The two values are compatible within the uncertainty estimated
for each line (0.21 and 0.28\,dex, see Table\,\ref{errors}). The bottom panel of
Fig. \ref{heavy} shows that both values are compatible with the Galactic trend,
with the lower value given by the YII line at 509\,nm giving a better match. 

The upper panel in Fig.\,\ref{heavy} presents the [Ba/Y] versus [Fe/H] abundance
ratio of star \#107 in comparison with the Galactic trend. We again plot the
abundance ratio, considering separately the abundances obtained for the two Y
lines.  This ratio is commonly used to discriminate the trend observed in
Galactic stars with respect to that observed in stars born in Milky Way
satellite galaxies \citep[][]{sbordone15}. E3 follows the trend observed in
Galactic stars.

\section{Summary and conclusions}\label{con}

Based on high-resolution spectra obtained with FLAMES-UVES/VLT, we presented a
chemical abundance analysis of four stars that are  likely members of the
globular cluster E3. The mean heliocentric cluster radial velocity and
metallicity are v$_{helio}$=12.6$\pm$0.4\,kms ($\sigma$=0.6$\pm$0.2\,\kms) and
[Fe/H]=-0.89$\pm$0.08\,dex. This RV is consistent with the 8.9$\pm$4\,\kms\
reported by SS15 from low-resolution spectroscopy. The iron content is in good
agreement with the FM15 value, which is however deduced from the spectroscopy of
two stars whose RV is significantly different from our values, namely
45$\pm$5\,\kms.

We have presented abundances of the light element Na, iron-peak (Fe, V, Cr, Ni,
Mn), $\alpha$ (Mg, Si, Ca, Ti), and neutron-capture (Y, Ba, Eu) elements. Mn, Y,
and Eu abundances were only measured for star \#107, having the highest S/N
spectrum. E3 has abundance ratios of its $\alpha$, iron-peak, and
neutron-capture elements that are fully compatible with the trends observed in
Galactic stars, and its $\alpha$-enhancement is typical (0.3--0.4\,dex) for a
halo GC.

We did not detect any significant spread in  Na abundance. The mean cluster Na
abundance is [Na/Fe]=0.18$\pm$0.07\,dex, with a range between 0.12 and
0.29\,dex, i.e., less than 0.2\,dex. This mean value and observed spread place
E3 in the portion of the Na-O diagram (Fig. \ref{na2}) occupied mainly by FG
stars and is close to the separation between FG and SG stars for the similar
metallicity clusters M4 and 47\,Tuc and the intermediate- and low-mass clusters
NGC\,288 and NGC\,6535. The Na abundance of E3 stars is also similar to FG stars
in the low-mass cluster Ter\,8.

With a luminosity and mass slightly lower than E3 \citep[M$_V$=-4.12,
M=1.4$\times$10$^4$M$_\odot$, SS15,][]{harris96}, \citet[][]{simpson17} detected
evidence of multiple populations in the GC ESO452-SC11 (M$_V$=-4.02,
M=6.8$\times$10$^3$M$_\odot$). They suggest that most of the stars they observed
in this cluster, appear to belong to the SG based on their high CN index
strengths. Indeed, E3 stars appear to show low CN index strength compared with
NGC1851 and ESO452-SC11 \citep[see Fig.11 of][]{simpson17}. 

We conclude that E3 is dominated by FG stars, consistently with what is expected
on the basis of its low luminosity \citep[][]{milone17}. Our results support the
SS15 conclusion that E3 may be a single stellar population cluster. The easiest
interpretation  suggests that the low mass of the cluster did not allow  the
polluters from which SG stars would have formed to be retained. However, our
limited statistics do not allow us to exclude the presence of a minor component
of SG stars.

While mass is commonly considered an important factor for the development of
MPs, others have been found to be relevant, such as  age and metallicity and
probably the environment \citep[][]{carretta10,bastian17}. In addition to E3, a
few more low-mass GCs are known that do not  present evidence of MPs. Pal12 and
Ter7 \citep[M$_V$=-4.47, -5.01,][]{cohen04,sbordone07} are known to belong to
the Sagittarius dwarf spheroidal galaxy (Sgr dSph) or to have once been part of
this galaxy. They are also known to be younger than the bulk of the Galactic GCs
\citep[][]{marinfranch09}. The old, low-mass Sgr cluster Ter\,8  (M$_V$=-5.07)
also hosts a minority  of SG stars  and is dominated by FG stars
\citep[][]{carretta14}. Rup\,106, similarly to E3 (FM15), has an age typical of
the bulk of Galactic GCs, but it has also been suggested that it has  abundance
patterns pointing towards an extragalactic origin \citep[][]{villanova13}. It is
currently the most massive (M$_V$=-6.35) Galactic GC known not to host MPs. On
the other hand, a number of relatively massive (M$\simeq$10$^5$M$_\odot$)
clusters younger than about 2\,Gyr have been found that do not  host MPs
\citep[][]{mucciarelli08,mucciarelli14,martocchia17,bastian17,martocchia18}.

The body of evidence accumulated so far supports E3 as a globular cluster
dominated by first-generation stars, and having age and chemical composition
similar to the bulk of the Galactic GC population.

\begin{acknowledgements}

We thank the anonymous referee and the editor Eline Tolstoy for a careful
reading of the paper and constructive comments which improved the quality of the
presentation. L.M. acknowledges support from ``Proyecto Interno'' of the
Universidad Andres Bello. S.V. and C.M.B. gratefully acknowledge the support
provided by FONDECYT N.1170518 and 1150060, respectively. This research has made
use of the NASA Astrophysics Data System and of the SIMBAD database, operated at
CDS, Strasbourg, France. This work has made use of data from the European Space
Agency (ESA) mission {\it Gaia} (\url{https://www.cosmos.esa.int/gaia}),
processed by the {\it Gaia} Data Processing and Analysis Consortium (DPAC,
\url{https://www.cosmos.esa.int/web/gaia/dpac/consortium}). Funding for the DPAC
has been provided by national institutions, in particular the institutions
participating in the {\it Gaia} Multilateral Agreement.

\end{acknowledgements}

%
%

\end{document}